# Magnetically Enhanced Thermoelectric Effect Driven by Martensitic Transformation in the Weak Itinerant Ferromagnet Co$_2$NbSn

Takumi Kihara[1]*, Xiao Xu[2], Yuki Ogi[3], Yoshiya Adachi[3], Tufan Roy[4, 5], Ryuji Matsuura[6], and Takeshi Kanomata[7]

[1] *Research Institute for Interdisciplinary Science, Okayama University, Okayama, Okayama 700-8530, Japan*
[2] *Department of Materials Science, Tohoku University, Sendai, Miyagi 980-8579, Japan*
[3] *Graduate School of Science and Engineering, Yamagata University, Yonezawa, Yamagata 990-8560, Japan*
[4] *Center for Science and Innovation in Spintronics (CSIS), Core Research Cluster (CRC), Tohoku University, Sendai, Miyagi 980-8577, Japan*
[5] *Research Institute of Electrical Communication (RIEC), Tohoku University, Sendai, Miyagi 980-8577, Japan*
[6] *Faculty of Engineering, Tohoku Gakuin University, Sendai, Miyagi 984-8588, Japan*
[7] *Research Institute for Engineering and Technology, Tohoku Gakuin University, Sendai, Miyagi 984-8588, Japan*

We investigated the magnetic and thermoelectric properties of the full Heusler alloy Co$_2$NbSn, which exhibits a martensitic transformation at 240 K. Magnetization measurements reveal weak itinerant ferromagnetism in the martensitic phase, which is well described by Takahashi's spin fluctuation theory. The characteristic spin fluctuation parameters were estimated to be $T_0 = 1.0 \times 10^3$ K and $T_A = 7.2 \times 10^3$ K. Seebeck coefficient measurements under magnetic fields up to 9 T show complex temperature and field dependence, which we decomposed into electron diffusion, spin fluctuation drag, and magnon drag components. A significant magnon-drag contribution was identified in both austenite and martensitic phases. Remarkably, this contribution is strongly enhanced in the martensitic phase compared to the austenite phase, despite a smaller magnetic moment. These findings provide evidence for robust low-energy spin excitations and highlight the potential of martensitic transformation in enhancing the thermoelectric performance of itinerant ferromagnetic alloys.

Thermoelectric (TE) effects are key phenomena that enable the direct and reversible conversion between heat and electricity[1]. The development of high-performance TE materials has traditionally focused on engineering electron and phonon transport. However, this approach has reached a performance plateau due to the intrinsic trade-off between electrical and thermal conductivities. To overcome this limitation, incorporating additional degrees of freedom, such as spin, has recently emerged as a promising strategy. Spin-driven thermoelectric phenomena including magnon-drag, spin-fluctuation-drag, and spin entropy contributions can provide additional channels for electric charge and entropy transport, leading to enhanced thermopower and power factor in magnetic materials[2–6]. In particular, itinerant ferromagnetic and antiferromagnetic materials have shown notable thermoelectric enhancements driven by spin excitations[4,5,7]. However, theoretical models explaining such spin-driven thermoelectric effects, including their temperature and magnetic field dependences, remain limited, partly due to the lack of experimental studies that comprehensively investigate both magnetism and TE properties.

Among spin-driven TE materials, Co-based Heusler alloys have attracted increasing attention owing to their pronounced spin-fluctuation-drag and magnon-drag effects[5]. In this study, we focus on the weak itinerant ferromagnetic $Co_2NbSn$, which undergoes a martensitic transformation (MT) at 240 K from the cubic austenite (A) phase to a lower-symmetry martensitic (M) phase. This transition enables us to investigate the influence of MT on TE properties. Through systematic measurements of magnetic and TE behavior, we decomposed the Seebeck coefficient into electron diffusion, spin-fluctuation drag, and magnon-drag components. Notably, the M phase exhibits an enhanced magnon-drag contribution despite a smaller magnetic moment, providing compelling evidence for strong low-energy magnetic excitations associated with the MT.

Polycrystalline samples of $Co_2NbSn$ were synthesized using arc melting. The ingots were vacuum-sealed in a quartz tube and annealed at 800 °C for 3 days, followed by an additional annealing at 450 °C for 3 days to achieve high compositional homogeneity, and then quenched in cold water. Temperature and magnetic field variations of magnetization were taken using a Quantum Design MPMS. Electric and thermal transport, as well as the Seebeck effect, were simultaneously measured for the rectangular prism-shaped polycrystalline sample with dimensions of $12.9 \times 1.2 \times 1.0$ mm$^3$, employing the Thermal Transport Option (TTO) in a Quantum Design PPMS.

First-principles calculations of the electronic, structural, and magnetic properties were carried out using the vienna ab initio simulation package (VASP)[8,9] with the projector augmented wave method[10]. A plane-wave cutoff energy of 500 eV and a $16 \times 16 \times 16$ $k$-mesh were employed for self-consistent-field calculation. The generalized gradient approximation (GGA) was used for the exchange correlation functional[11].

Magnetization measurements were performed between 5 K and 200 K and magnetic field up to 7 T, as shown in Figs. 1(a)-1(f). Since Co$_2$NbSn undergoes a MT from cubic A phase to orthorhombic M phase at $T_{\text{MT}} = 240$ K with decreasing temperature[12–14], the presented data reflect the properties of the M phase. As shown in Fig. 1(a), the $M-T$ curves measured at 0.01, 0.1, and 1 T exhibit a ferromagnetic phase transition. The Curie temperature of the M phase, $T_{\text{C}}^{\text{M}} = 109.6$ K, was determined from the dip in the temperature derivative of the magnetization at 0.01 T, as indicated by the arrows in Fig. 1(b). By fitting the paramagnetic region of the $M-T$ curve at 1 T to the modified Curie-Weiss law [$\chi = M(T)/\mu_0 H = C/(T-T_\theta) + \chi_0$], the effective magnetic moment $p_{\text{eff}} = g\sqrt{S(S+1)} \simeq 1.81$ $\mu_{\text{B}}$/Co was obtained, where $C = N p_{\text{eff}}^2 \mu_{\text{B}}^2 / 3k_{\text{B}}$ and $g=2$. $M-H$ curves measured at various temperatures are shown in Fig. 1(c). As shown as black solid curve in Fig. 1(c), the data at 5 K shows no saturation at 7 T, with a finite slope of $dM/\mu_0 dH \sim 0.081$ Am$^2$/kg·T, consistent with itinerant ferromagnetism[15]. Figure 1(d) represents Arrott plots ($M^2$ vs. $H/M$) with linear fits to the high field regime. Spontaneous magnetization $M_{\text{s}}(T)$ was estimated by extrapolating to $H/M = 0$. The obtained $M_{\text{s}}(5\text{ K}) \sim 0.38$ $\mu_{\text{B}}$/Co-atom is slightly larger than the previously reported values[12–14,16], indicating improved sample homogeneity due to the annealing process.

We employed Takahashi's spin fluctuation theory[17,18] to analyze the itinerant electron magnetism of Co$_2$NbSn, where the total amplitude of the local spin fluctuation $\langle S_{\text{L}}^2 \rangle_{\text{tot}}$ comprises the thermal $\langle S_{\text{L}}^2 \rangle_{\text{T}}$ and the zero point (quantum) fluctuations $\langle S_{\text{L}}^2 \rangle_{\text{Q}}$ with opposing temperature dependence. Consequently, $\langle S_{\text{L}}^2 \rangle_{\text{tot}} = \langle S_{\text{L}}^2 \rangle_{\text{T}} + \langle S_{\text{L}}^2 \rangle_{\text{Q}}$ is temperature independent, and therefore, it was assumed to be constant in Takahashi's spin fluctuation theory. The validity of this theory was demonstrated in numerous experimental studies on 3$d$ transition metal intermetallic compounds[19–23], Heusler alloys[15,24–28], and 4$f$/5$f$ electron systems[29,30]. In the ferromagnetic ground state, the assumption of $\langle S_{\text{L}}^2 \rangle_{\text{tot}}$ conservation leads to

$$H = \frac{2T_{\text{A}}^2}{15cN_0^3(g\mu_{\text{B}})^4 T_0}[-M_{\text{s}}^2(0) + M^2]M, \qquad (1)$$

where $g$ represents the Lande' $g$ factor, $c = 1/2$, $N_0$ is the number of magnetic sites, $\mu_B$ is the Bohr magneton, and $M_s(0)$ is the spontaneous magnetization at 0 K. The spectral parameters $T_A$ and $T_0$, representing the spectral width in the wave vector and energy space, respectively, characterize the itinerant magnetic nature of a material. By modifying this equation as follows, it can be directly compared with the experimentally obtainable Arrott plot.

$$\frac{H}{M} = a(0) + b(0)M^2,$$

$$a(0) = -\frac{M_0^2}{b(0)}, \quad b(0) = \frac{F_1}{N_0^3(g\mu_B)^4}, \quad (2)$$

$$F_1 = \frac{2T_A^2}{15cT_0}.$$

The mode-mode coupling term $F_1$ was estimated to be $F_1 = 1.3 \times 10^3$ K from the slope of the Arrott plot at 5 K [Fig. 1(d)]. $T_0$ and $T_A$ are estimated using $F_1$ through following relations:

$$\left(\frac{T_C}{T_0}\right)^{5/6} = \frac{p_s^2}{5g^2 C_{4/3}} \left(\frac{15cF_1}{2T_C}\right)^{1/2}, \quad (3)$$

$$\left(\frac{T_C}{T_A}\right)^{5/3} = \frac{p_s^2}{5g^2 C_{4/3}} \left(\frac{2T_C}{15cF_1}\right)^{1/3}, \quad (4)$$

where $C_{4/3} = 1.00608\cdots$ and $p_s$ is the spontaneous magnetic moment ($p_s = M_s(0)/N_0\mu_B$). Using Eqs. (2)-(4), we obtained $T_0 = 1.0 \times 10^3$ K and $T_A = 7.2 \times 10^3$ K for Co$_2$NbSn, respectively, which are comparable to those of other Co based Heusler alloys such as Co$_2$NbGa[15], Co$_2$TiGa[31], and Co$_2$ZrAl[32]. An important consequence from Takahashi's spin fluctuation theory is the linear temperature dependence of the squared spontaneous magnetization and the linear $H/M$ dependence of the $M^4$ at $T_C$, as expressed in Eqs. (5), (6).

$$\left[\frac{p_s(T)}{p_s(0)}\right]^2 = 1 - \frac{112.1}{p_s^4(0)} \left(\frac{T}{T_A}\right)^2 \quad (5)$$

$$\frac{H}{M} = \frac{k_B T_A^3}{2[3\pi T_C(2+\sqrt{5})^2]} \cdot \frac{M^4}{N_0^5 \mu_B^6} \quad (6)$$

Here, $k_B$ denotes Boltzmann constant. As demonstrated in Figs. 1(e) and 1(f), the data for Co$_2$NbSn clearly exhibits these relationships. Furthermore, the estimated values of $T_A$, $8.6 \times 10^3$ K from Eq. (5) and $7.6 \times 10^3$ K from Eq. (6) are consistent with the value determined using the data at 5 K. It is noteworthy that Takahashi's spin fluctuation theory demonstrates the itinerant ferromagnetic nature of Co$_2$NbSn not only at around ground state, but also across a wide temperature range up to $T_C$. This framework also enables systematic evaluation of spin fluctuation contributions in various materials[18,19,29,30,33,34]. Figure 2 shows log-log plot of $p_{\text{eff}}/p_{\text{s}}$ versus $T_C/T_0$ for Co$_2$NbSn, other Co-based and NiMn-based Heusler alloys, and pure Ni. The dashed line represents the theoretical universal relation: $p_{\text{eff}}/p_{\text{s}} \simeq 1.4(T_C/T_0)^{-2/3}$. In this framework, localized moment systems are concentrated around $p_{\text{eff}}/p_{\text{s}} = T_C/T_0 \sim 1$, because $T_C \sim T_0$. In contrast, in the Rhodes-Wohlfarth plot (i.e., $p_C/p_S$ versus $T_C$, where $p_{\text{eff}} = \sqrt{p_C(p_C + 2)}$), these systems are spread over a wide range of $T_C$ with $p_C/p_s \sim 1$ [35,36]. While NiMn-based alloys are close to the localized ferromagnets, Co-based alloys, including Co$_2$NbSn with $p_{\text{eff}}/p_{\text{s}} \gg 1$, are classified into weak ferromagnets as shown in Fig. 2. We also plotted Fe$_2$V$_{0.9}$Cr$_{0.1}$Al$_{0.9}$Si$_{0.1}$ using $p_{\text{eff}}/p_{\text{s}} = 8.2$[4] and the above theoretical relation, and estimated $T_0 \sim 2.3 \times 10^3$ K using $T_C = 160$ K, which is consistent with its weak itinerant ferromagnetic nature[4,37].

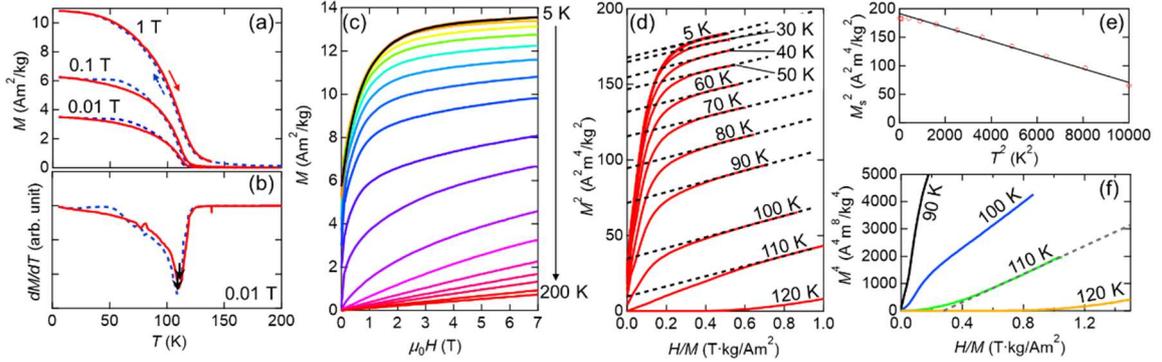

Fig. 1. (Color online) Temperature and magnetic field dependence of the magnetization for Co$_2$NbSn: (a) $M - T$ curves measured at 0.01, 0.1, and 1 T. (b) Temperature derivative of the magnetization $dM/dT$ at 0.01 T. (c) $M - H$ curves measured up to 7 T at the various temperatures below 200 K. (d) Arrott plots ($M^2$ vs. $H/M$) with dotted lines representing the fit to the high-field regime. (e) Squared spontaneous magnetization $M_s^2$ as a function of squared temperature $T^2$. (f) $M^4$ vs. $H/M$ plots with dotted lines representing the fit to the high-field regime.

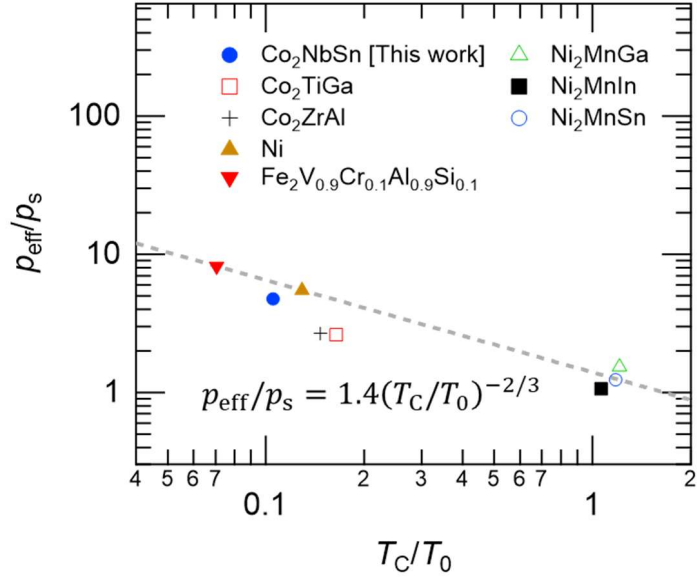

Fig. 2 (Color online) Double-logarithmic plot of $p_{\text{eff}}/p_s$ versus $T_C/T_0$ for $Co_2NbSn$, other Co-based and NiMn-based Heusler alloys, $Fe_2V_{0.9}Cr_{0.1}Al_{0.9}Si_{0.1}$, and pure Ni.

Figure 3(a) shows the temperature dependence of electrical resistivity measured under magnetic fields up to 9 T. A clear jump is observed around $T_{\text{MT}} = 240$ K for all measured fields, which corresponds to the MT. Above $T_{\text{MT}}$, the resistivity exhibits a linear temperature dependence, characteristic of the metallic A phase. Between 110 K and 240 K, it becomes nearly constant, indicating strong magnetic scattering in the paramagnetic M phase. Below $T_C^M$, resistivity decreases with decreasing temperature due to the suppression of magnetic scattering by the onset of ferromagnetic ordering. The thermal conductivity exhibits a kink at $T_{\text{MT}}$, and shows a similar trend to resistivity in the M phase [Fig. 3(b)]. Figure 3(c) shows the temperature dependence of the Seebeck coefficient $S(T)$ measured under magnetic fields ranging from 0 to 9 T. Throughout the entire temperature range below room temperature, the negative values are observed, indicating that electrons are the dominant carriers. Above $T_{\text{MT}}$, $S(T)$ exhibits a linear temperature dependence in the paramagnetic A phase, which is attributable to the well-known electron diffusion term ($S_d$). $S_d$ of a free electron gas can be expressed as:

$$S_d = -\frac{\pi^2}{3}\frac{k_B^2 T}{e}\frac{1}{D(E_F)}\left(\frac{\partial D(E)}{\partial E}\right)_{E=E_F},$$

where $e$, $D(E)$, and $E_F$ denote electron charge, density of states (DOS), and fermi energy, respectively[1,38]. Through linear fitting of $S(T)$ above $T_{\text{MT}}$, a slope of approximately -0.017 μV/K² is obtained, which is nearly consistent with the value estimated from the theoretically calculated DOS for the nonmagnetic A phase[39]. In contrast, the thermoelectric behavior in M phase is more complex. The absolute value of $S(T)$ increases sharply at $T_{\mathrm{MT}}$ upon cooling, reaches a peak around 130 K, and then decreases to zero with further decreasing temperature. The maximum absolute value of $S(T)$ reaches 21 μV/K, which is comparable to that of the A phase in other Co-based[5,40] and NiMn-based[41,42] Heusler alloys. The application of magnetic fields suppresses $|S(T)|$. Figure 3(d) shows the Seebeck coefficients from 0 to 7 T, obtained by subtracting the 9 T data. Their absolute values exhibit a clear peak around 70 K, as indicated by $T_{\mathrm{peak}}^{\mathrm{sf}}$ in Fig. 3(d). These magnetic field dependent components of the Seebeck coefficients suggest the presence of a magnetically induced contribution to the thermoelectric response. Similar behavior has been reported for Fe$_2$VAl-based[4] and Co$_2$XAl (X = Ti, V, and Nb)[5] Heusler alloys, and has been interpreted as a manifestation of the spin-fluctuation-drag effect[43]. Accordingly, we denote these field dependent contributions as the spin fluctuation component $S_{\mathrm{sf}}$. It is noteworthy that $T_{\mathrm{peak}}^{\mathrm{sf}}$ is significantly lower than $T_{\mathrm{C}}^{\mathrm{M}}$, whereas these temperatures are nearly identical in previous reports on Fe$_2$VAl-based[4] and Co$_2$XAl (X = Ti, V, and Nb)[5] Heusler alloys. This indicates the intensity of low-energy spin fluctuations reaches a maximum below $T_{\mathrm{C}}^{\mathrm{M}}$. This behavior cannot be explained solely by the thermal spin fluctuation, since $\langle S_{\mathrm{L}}^2 \rangle_{\mathrm{T}}$ increases monotonically with increasing temperature below $T_{\mathrm{C}}^{\mathrm{M}}$. Therefore, it suggests the existence of a strong intensity of low-energy spin fluctuations, which decreases with increasing temperature below $T_{\mathrm{C}}^{\mathrm{M}}$. As mentioned above, according to Takahashi's spin fluctuation theory, the total spin fluctuation consists of thermal and quantum components, which exhibit opposite temperature dependences. Consequently, the energy and wave-vector distributions of the total spin fluctuation, as well as its temperature evolution can be highly complex. The deviation of $T_{\mathrm{peak}}^{\mathrm{sf}}$ from $T_{\mathrm{C}}^{\mathrm{M}}$ in the M phase of Co$_2$NbSn may indicate that, in addition to the thermal component, the low-energy part of quantum spin fluctuations plays an essential role. In contrast, only the thermal component appears to contribute to $S_{\mathrm{sf}}$ in the A phase of Fe$_2$VAl-based and Co$_2$XAl (X = Ti, V, and Nb) alloys[4,5].

The spin-fluctuation-drag effect in the theoretical model proposed by Okabe also exhibits a peak structure and successfully explains the low temperature behavior of $S(T)$ observed in $A$Fe$_4$Sb$_{12}$ ($A$ = Ba, Sr, and Ca)[43,44]. In this model, narrow-band ferromagnetic electrons are responsible for spin fluctuations, while wide-band nonmagnetic electrons contribute to the electrical conductivity. The peak of $|S(T)|$ arises from a crossover between the spin-fluctuation-drag-dominant regime and the electron-diffusion-dominant regime. However, the spin fluctuation component $S_{\text{sf}}$, shown in Fig. 3(d) for Co$_2$NbSn, can be well reproduced only when the system is assumed to be hole-like, which contradicts the experimental results shown in Fig. 3(c)[39]. Therefore, we conclude that this model cannot adequately explain the temperature dependence of $S_{\text{sf}}$ in the M phase of Co$_2$NbSn.

The temperature dependences of the power factor (PF; $S^2/\rho$) and the dimensionless figure of merit [$ZT = S^2T/(\rho\kappa)$] are shown in Figs. 3(e) and 3(f), respectively. Both PF and $ZT$ exhibit a peak within the paramagnetic M phase below 240 K, reflecting the magnetic contributions to $\rho$, $\kappa$, and $S$. It is noteworthy that the maximum value of PF at 148 K reaches 161 $\mu$W·m$^{-1}$·K$^{-2}$, which is approximately 60% higher than that of the A phase just above $T_{\text{MT}}$. As discussed below, this enhancement of the PF is attributed to the magnon-drag effect in M phase.

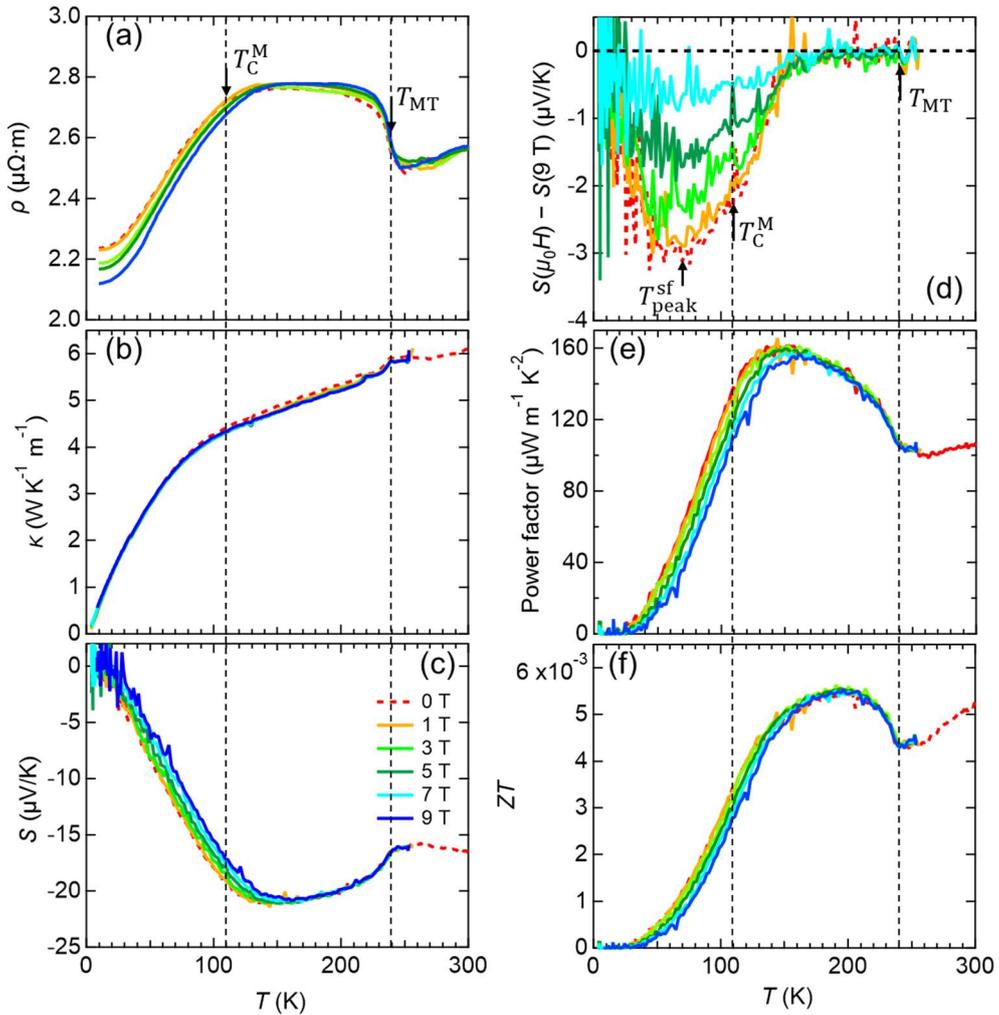

Fig. 3 (Color online) Temperature dependence of (a) electrical resistivity, (b)

thermal conductivity, (c) Seebeck coefficient, (e) power factor, and (f) figure of merit $ZT$, measured at various magnetic fields up to 9 T. Panel (d) shows the Seebeck coefficients after subtracting the data measured at 9 T.

To further analyze the origin of the magnetic contribution to the Seebeck effect in $Co_2NbSn$, we estimated its temperature dependence by subtracting the electron diffusion term $S_d$ from the experimental data at 0 T. Here, since the DOS and its energy derivative for ferromagnetic M phase are close to those for the nonmagnetic A phase (see Fig. S2 in the Supplemental information[39]), we used $S_d = -0.017T$ as the electron diffusion term for the M phase. After subtracting the $S_d$ and $S_{sf}$ terms from $S_{tot}$, which corresponds to the experimental data at 0 T, a large Seebeck coefficient remains. This remaining component is attributed to magnon drag effect, as discussed below, and is therefore denoted as $S_{md}$, where $S_{tot} = S_d + S_{sf} + S_{md}$. Figure 4 shows the temperature dependences of the $S_d$, $S_{sf}$, and $S_{md}$ components along with the total Seebeck coefficient $S_{tot}$. The absolute value of $S_{md}$ increases with increasing temperature and reaches a maximum at $T_{peak}^{md} = 153$ K, which is approximately 40 % higher than $T_C^M$. Above this temperature, it decreases upon further heating up to $T_{MT}$, and becomes nearly temperature independent above $T_{MT}$. As shown in Fig. S4 in the Supplemental Information[39], the temperature dependence of $|S_{md}|$ below $T_C^M$ can be well fitted by a $T^{3/2}$ function, which is characteristic of the magnon-drag effect in the case of quadratic magnon dispersion. This behavior is described by the relation $S_{md} = \pm \frac{2}{3}\frac{C_m}{n_e e}\frac{1}{1+\tau_{em}/\tau_m}$, where $n_e$ and $C_m$ are the carrier density and the magnon specific heat, respectively. $\tau_m$ and $\tau_{em}$ denote the relaxation times for magnon-magnon and magnon-electron scattering, respectively. For a ferromagnetic system, $C_m \propto T^{3/2}$; therefore, the estimated $S_{md}$ can be interpreted as the contribution from the magnon-drag effect, provided that $\tau_m$ and $\tau_{em}$ are temperature-independent. The temperature-insensitive component observed in the paramagnetic A phase can be interpreted as the so-called paramagnon-drag effect, which has been observed in $Co_2XAl$ ($X$ = Ti, V, and Nb) Heusler alloys[5] and Li-doped MnTe[7]. In these materials, $|S_{md}|$ increases with temperature up to the Curie temperature and then saturates. This behavior is attributed to the persistence of short-range ferromagnetic correlations above the Curie temperature, which give rise to discrete magnon-like excitations and sustain the magnon-drag effect even in the paramagnetic phase. Therefore, the observed temperature-insensitive component suggests that short-range ferromagnetic order persists in the paramagnetic A phase at least up to room temperature. In contrast, $|S_{\mathrm{md}}|$ exhibits a negative temperature dependence between $T_{\mathrm{peak}}^{\mathrm{md}}$ and $T_{\mathrm{MT}}$, suggesting that such short-range ferromagnetic order is weakened and does not persist robustly in the paramagnetic M phase. It is noteworthy that $|S_{\mathrm{md}}| \sim 18.3$ $\mu$V/K at $T_{\mathrm{peak}}^{\mathrm{md}}$ is approximately 60% larger than that in the paramagnetic A phase, where the signal is attributed to the paramagnon-drag effect, despite the fact that the magnetic moment in M phase is smaller than that in A phase[14]. To the best of our knowledge, such a significant enhancement of the magnon-drag contribution to the Seebeck effect, induced by the MT, has been demonstrated for the first time. The observed discontinuous drop in $|S_{\mathrm{md}}|$ at $T_{\mathrm{MT}}$ likely reflects a reconstruction of the spin excitation spectrum and a concomitant reduction in short-range ferromagnetic correlations.

In summary, we have investigated the magnetic and thermoelectric properties of $Co_2NbSn$, which undergoes a martensitic transformation at 240 K. Magnetization measurements revealed weak itinerant ferromagnetism in the martensitic phase, consistent with Takahashi's spin fluctuation theory. Seebeck coefficient measurements under magnetic fields up to 9 T exhibited clear contributions from spin-fluctuation-drag and magnon-drag effects. Notably, the spin-fluctuation-drag component exhibits a peak around 70 K, significantly below the Curie temperature, implying a key role of low-energy quantum spin fluctuations. Additionally, we demonstrated a pronounced enhancement of the magnon-drag effect via the martensitic transformation, indicating the presence of strong low-energy magnetic excitations and their essential role in enhancing the Seebeck effect.

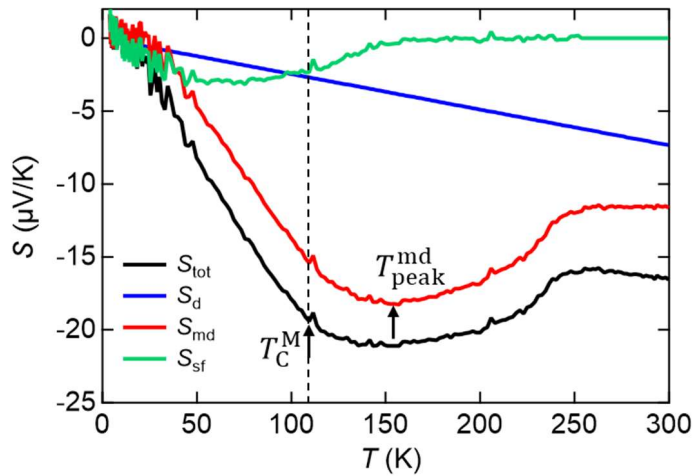

Fig. 4 (Color online) Temperature dependence of the electron diffusion component ($S_{\mathrm{d}}$), spin-fluctuation component ($S_{\mathrm{sf}}$), and magnon-drag component ($S_{\mathrm{md}}$),

plotted together with the total Seebeck coefficient ($S_\mathrm{tot}$) obtained at 0 T.

**Acknowledgment**

The authors are grateful to Professor Mitsuo Kataoka of Laboratory for Solid State Physics for his valuable contributions to the discussions. This work received partial support from the MEXT Leading Initiative for Excellent Young Researchers (Grant No. JPMXS0320210045), JSPS KAKENHI Grant-in-Aid for Scientific Research (B) (Grant No. JP23H01672), as well as the Murata Science Foundation, Sumitomo Foundation, and Okayama Foundation for Science and Technology.

*Address all correspondence to: tkihara@okayama-u.ac.jp

# Supplemental material for Magnetically Enhanced Thermoelectric Effect Driven by Martensitic Transformation in the Weak Itinerant Ferromagnet $Co_2NbSn$

Takumi Kihara[1]*, Xiao Xu[2], Yuki Ogi[3], Yoshiya Adachi[3], Tufan Roy[4,5], Ryuji Matsuura[6], and Takeshi Kanomata[7]

[1] Research Institute for Interdisciplinary Science, Okayama University, Okayama, Okayama 700-8530, Japan

[2] Department of Materials Science, Tohoku University, Sendai, Miyagi 980-8579, Japan

[3] Graduate School of Science and Engineering, Yamagata University, Yonezawa, Yamagata 990-8560, Japan

[4] Center for Science and Innovation in Spintronics (CSIS), Core Research Cluster (CRC), Tohoku University, Sendai, Miyagi 980-8577, Japan

[5] Research Institute of Electrical Communication (RIEC), Tohoku University, Sendai, Miyagi 980-8577, Japan

[6] Faculty of Engineering, Tohoku Gakuin University, Sendai, Miyagi 984-8588, Japan

[7] Research Institute for Engineering and Technology, Tohoku Gakuin University, Sendai, Miyagi 984-8588, Japan

*Address all correspondence to: tkihara@okayama-u.ac.jp

Figure S1 (a) shows the spin dependent band dispersion and the density of states (DOS) for the ferromagnetic austenite phase (cubic $L2_1$). At the Fermi level, an almost flat band appears in the minority spin channel, resulting in a large DOS at the Fermi level. Figure S1(b) shows the total energy as a function of the lattice constant ratio $c/a$. While the cubic state at $c/a = 1$ corresponds to a metastable state, the global energy minimum is observed in the tetragonal phase at $c/a = 1.12$, indicating occurrence of a martensitic transformation.

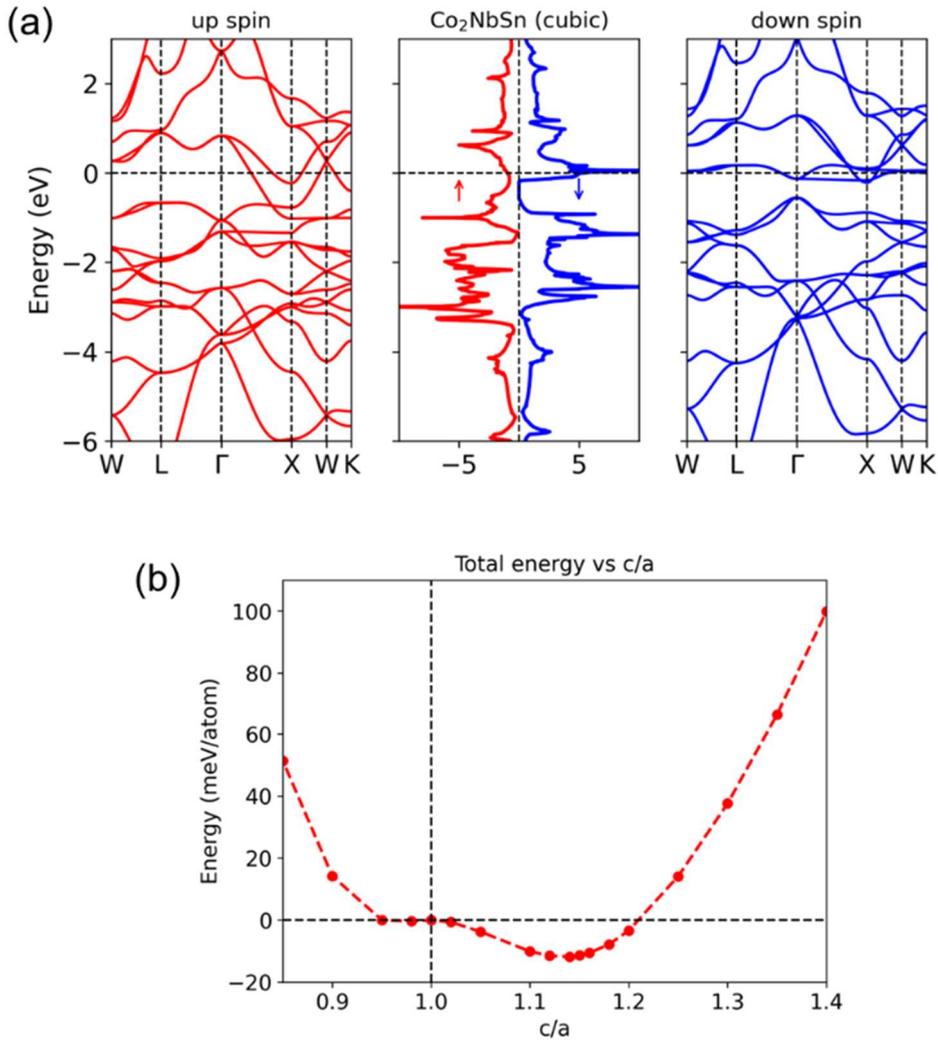

Fig. S1 (a) Spin dependent band structures (left and right) and density of states (center) of ferromagnetic austenite phase of $Co_2NbSn$, and (b) total energy as a function of lattice constant ratio $c/a$ for $Co_2NbSn$.

Figures S2(a) and S2(b) show the total DOS calculated for the ferromagnetic (FM) and nonmagnetic (NM) austenite phase, and the ferromagnetic martensitic phase. The DOS and its energy derivative at the Fermi level for FM martensitic phase are similar to those for NM austenite phase. Based on these results, the electron diffusion component of the Seebeck coefficient, $S_\text{d}$, is calculated using the relation $S_\text{d} = -\frac{\pi^2}{3}\frac{k_\text{B}^2 T}{e}\frac{1}{D(E_\text{F})}\left(\frac{\partial D(E)}{\partial E}\right)_{E=E_\text{F}}$. $S_\text{d} = -0.018T$ and $S_\text{d} = -0.024T$ are obtained for FM martensitic phase and NM austenite phase, respectively.

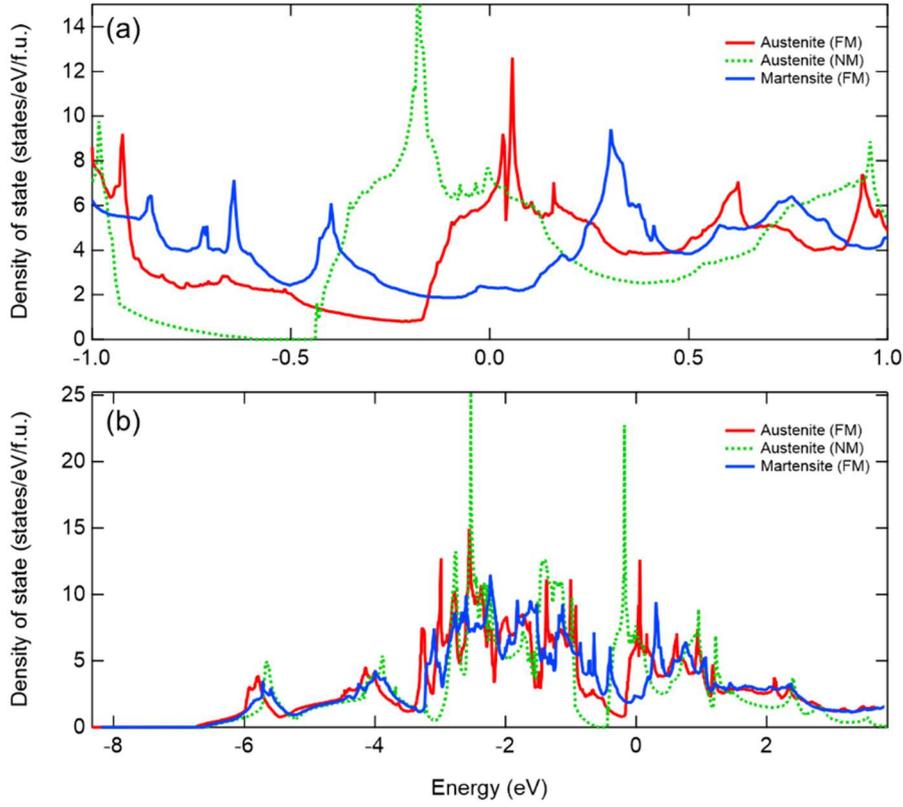

Fig. S2 Total density of states for the ferromagnetic (FM) and nonmagnetic (NM) austenite phases, and the ferromagnetic martensite phase. (a) Enlarged view around the Fermi energy $E = E_\text{F} = 0$. (b) Full energy range.

Figure S3 shows the temperature dependence of the spin-fluctuation component $S_{\text{sf}}$ of the Seebeck coefficient, obtained by subtracting the 9 T data from the 0 T data. The dotted curve represents a fit to the following function:

$$S(T) = \alpha T + \beta T \left(\frac{T}{\bar{T}}\right)^2 \log \frac{\delta + (T/\bar{T})^2}{(T/\bar{T})^2},$$

where $\alpha$, $\beta$, and $\bar{T}$ are the fitting parameters, and $\delta = 4$ (Ref. [42]). $\alpha = -0.0620$, $\beta = 4.08$, and $\bar{T} = 1860$ K are obtained. The experimental result can be well reproduced only when $\beta > 0$, indicating that the net carriers are hole-like. In addition, this component becomes nearly zero above 150 K, suggesting that the slope of the electron diffusion term remains unaffected by the applied magnetic field up to 9 T. Therefore, this extracted data shown in Fig. S3 is considered to exclude any contribution from the diffusion component, which can, in principle, be affected by magnetic field.

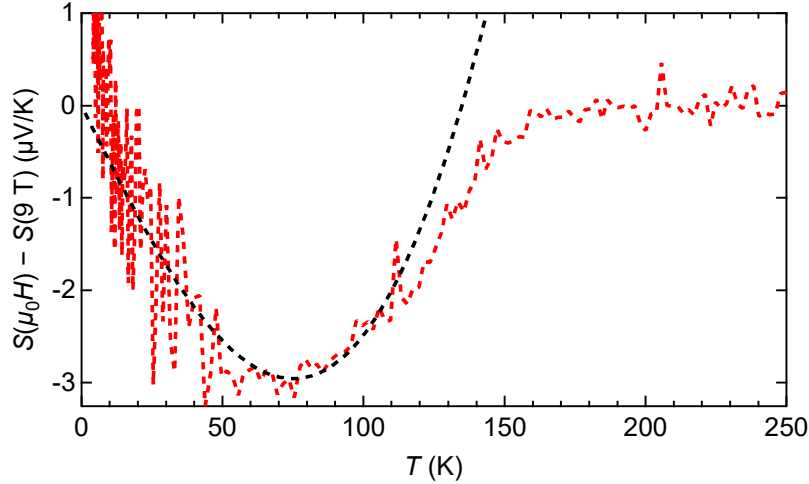

Fig. S3 Temperature dependence of the spin-fluctuation component $S_{\text{sf}}$ of the Seebeck coefficient. The dotted curve represents a fit to the function proposed in Ref. [42] in the main text.

Figure S4 shows a log-log plot of the absolute values of the total Seebeck coefficient $S_{\text{tot}}$, and the magnon-drag component $S_{\text{md}}$. The temperature dependence of $|S_{\text{md}}|$ below $T_C^M$ is approximately proportional to $T^{3/2}$, which corresponds to the temperature dependence of the magnon specific heat in a ferromagnet. This confirms that $S_{\text{md}}$ originates from the magnon-drag effect. In contrast, the temperature dependence of $|S_{\text{tot}}|$ deviates from the $T^{3/2}$ behavior because of additional contributions from the $S_{\text{d}}$ and $S_{\text{sf}}$ components. Consequently, the observation that only $S_{\text{md}}$ follows the $T^{3/2}$ dependence provides strong validation for the estimation of $S_{\text{d}}$ for the martensite phase.

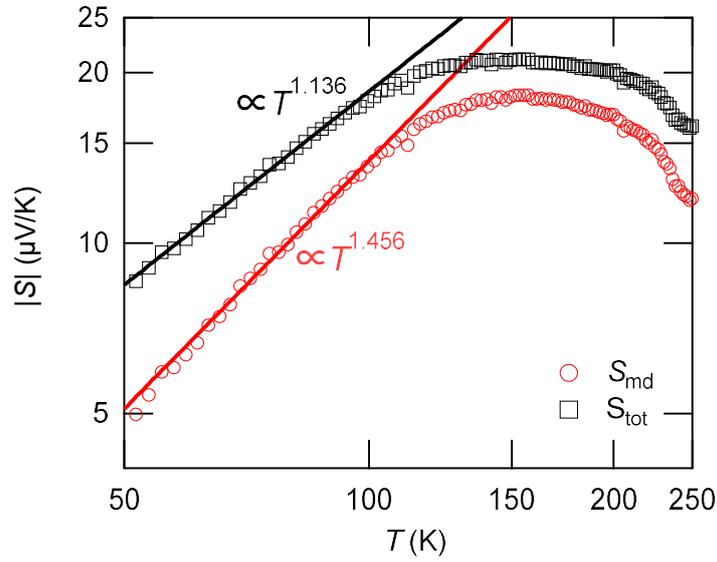

Fig. S4 Log-log plot of the absolute values of the total Seebeck coefficient $S_{\text{tot}}$, and the magnon-drag component $S_{\text{md}}$. The solid lines represent fits to a power-law function of temperature ($\propto T^n$).

18